\documentclass[12pt]{iopart}

\begin{document}
\title{Quantum mechanics and Leggett's principles of macroscopic realism}

\author{N L Chuprikov}

\address{Tomsk State Pedagogical University, 634041, Tomsk, Russia}

\begin{abstract}
On the basis of our recent model of a one-dimensional (1D) completed scattering we
argue that Leggett's principles of macroscopic realism must and can be extended onto
the level of single electrons and atoms. These principles need three quite feasible
innovations in quantum mechanics (QM): (1) at the conceptual level, QM must treat a
pure time-dependent one-particle state to involve two or more macroscopically distinct
alternatives for a particle as a pure {\it combined} one - the intermediate link
between a pure {\it elementary} state (indecomposable into macroscopically distinct
parts) and statistical mixture; (2) at the mathematical level, QM must provide the
presentation of the pure time-dependent combined state as a coherent superposition of
macroscopically distinct elementary states (MDESs); (3) at the experimental level, QM
must provide for such states two types of measurements - those for observing the
interference pattern resulting from the joint action of MDESs, and non-demolishing
'which-way' measurements for scanning the individual properties of MDESs.

\end{abstract}

\maketitle

\newcommand{\ppp}{\mbox{\hspace{5mm}}}
\newcommand{\ooo}{\mbox{\hspace{3mm}}}
\newcommand{\ooa}{\mbox{\hspace{1mm}}}

\section{Introduction}

At present one cannot imagine quantum mechanics (QM) without the Cat and EPR-Bell
paradoxes which have been in the focus of hot debates for a long time. As is known,
both the paradoxes are associated with the coherent superpositions of macroscopically
distinct states (CSMDSs), known as Cat states. So that, eventually, both they concern
the superposition principle and its role in the universe.

The aim of this paper is twofold: (\i) to show that the current status of the
superposition principle, which has been formed in modern physics on the basis of the
lessons learned from these two paradoxes, is controversial and needed in revision;
(\i\i) to present a new solution of the problem, based on our recent model of a
one-dimensional (1D) completed scattering.

\section{The puzzle of Cat states in quantum mechanics: two paradoxes - two
mutually exclusive lessons} \label{s1}

We begin our analysis with the Cat paradox, as namely here the notion of Cat states
was first introduced. As is known, the main participants of the paradox are a
radioactive nucleus, a vial of a poison gas and cat, all being in an isolated box. It
is suggested that just before opening the box the cat is died if the pial has been
broken; and, in its turn, the pial is broken if the nucleus has decayed. Otherwise,
the cat remains alive.

However, for our goals it is suitable to change this setting, with no distorting its
essence. Namely, let the role of the nucleus be played by an electron scattering on a
1D potential barrier, and let the cat be alive when the electron is reflected by the
barrier; otherwise, when the electron is transmitted, it is died. Then, according to
the usual practice of setting this thought experiment, as a quantum-mechanical
problem, we shall consider the electron and cat as parts of the compound system
'electron+cat' to be in a pure state presented in terms of the electron's and cat's
states.

For example, let $|\Psi_{tr}^{end}\rangle$ and $|\Psi_{ref}^{end}\rangle$ be final
pure states of a transmitted and reflected electron, respectively. Similarly, let
$|0\rangle_c$ and $|1\rangle_c$ be pure states of a died and alive cat, respectively.
Then a pure state $|\Psi\rangle_{e+c}$ of the 'electron+cat' system is
\begin{eqnarray} \label{1}
\fl |\Psi\rangle_{e+c}=c_0|0\rangle_{e+c}+c_1 |1\rangle_{e+c},
\end{eqnarray}
where $|c_0|^2+|c_1|^2=1$; $|0\rangle_{e+c}=|\Psi_{tr}^{end}\rangle\cdot|0\rangle_c$
and $|1\rangle_{e+c}=|\Psi_{ref}^{end}\rangle\cdot|1\rangle_c$.

The state (\ref{1}) is just the Cat state to represent a CSMDS. The present vision of
the Cat paradox involves at least three aspects of the problem associated with this
state.

(\i) {\it This paradox is treated \cite{Gh1,Gh2} as the macro-objectification
problem}. It shows that the superposition principle contradicts the principles of
macroscopic realism (PMRs) \cite{Le1,Le2,Le3} to govern the classical world. Indeed,
its current formulation implies that the state (\ref{1}) is an elementary one, i.e.,
it is indecomposable into parts (irrespective of whether or not its constituents
interfere with each other). It implies that namely the state $|\Psi\rangle_{e+c}$,
rather than $|0\rangle_{e+c}$ and $|1\rangle_{e+c}$, must be endowed with observables.
That is, in fact, the superposition principle demands the cat to be died and alive
simultaneously. Of course, such a requirement contradicts the PMRs. By them, the cat
must be in a definite state at any instant of time, and all physical observables can
be introduced for $|0\rangle_{e+c}$ and $|1\rangle_{e+c}$, rather than for
$|\Psi\rangle_{e+c}$.

(\i\i) {\it This paradox is often treated as a 'measurement' problem.} Indeed, in this
thought experiment the cat can be considered as the symbol of the pointer of a
macroscopic device to measure the final electron's state. From this viewpoint, the
state (\ref{1}) corresponds to a nonphysical situation when the pointer is not in a
definite state.

(\i\i\i) {\it This paradox leads to the problem of quantum entanglement and quantum
nonlocality.} Indeed, by Schr\"odinger the state (\ref{1}) is an entangled one. This
notion has been introduced by him in order to mark the novel type of relationship
between the participants of the paradox. It is irreducible to an interaction and in
fact inspired by the formula (\ref{1}). The nonlocal properties of such a relationship
have been revealed in the EPR-Bell paradox. As it has been shown theoretically and
experimentally, a combined system, being in a CSMDS, exhibits nonzero correlations
between two events to occur for its different parts separated by a spatial-like
interval. By the no-signalling theories, the above nonzero correlations does not mean
that these events can be linked with a causal signal. However, the concept of
influence without interaction is, perhaps, the most moot one in modern physics.
Rather, the superposition principle again conflicts with classical physics - now with
its principles of special relativity.

It is not difficult to see that, within the current interpretation of Cat states, the
Cat and EPR-Bell paradoxes give two cardinally different lessons about the universe
and QM. By them there are in fact two universes and two QMs. Indeed,

(1) "Cat lesson" teaches us that the universe to appear in our perception is local
(i.e., it respects the PMSs) and that the superposition principle and hence QM itself
do not govern this universe;

(2) "EPR-Bell lesson" says that the universe is nonlocal and QM possesses the
plenipotentiary power in this universe, because all the niceties of the Cat-state's
nonlocality are within the grasp of a classical device.

It is seen that these lessons exhibit exactly the opposite attitudes to the
relationship between quantum dynamics (to obey the superposition principle) and
classical devices (to respect the PMRs): the Cat lesson teaches us that any classical
devise cannot in principle to grasp the nonlocality of CSMDSs; however, the EPR-Bell
lesson appeals namely to it, as the highest instance, to confirm the reality of
quantum nonlocality.

It seems to have been impossible even on the mathematical level to respect, in one
theory, these two logically ambivalent lessons. However, such program has been
realized within the framework of the GWRP-approach (see \cite{Gh1,Gh2,Pe1} and the
references therein). By this model the universe comprises both the properties: it is
nonlocal at the level of single electrons and atoms, but (practically) local at the
level of macro-objects. The model is checkable experimentally (for deep analyses of
this and other approaches to the macro-objectification problem see
\cite{Le1,Le2,Le3,Gh1,Gh2,Pe1,Sc1,Sc2,Ad1,Ad2,Har,Omn,Ba1}) and is now considered as
the most prominent alternative quantum theory of the universe.

However, on the ontological level, a new model of the universe remains unclear in many
respects. For example, such terms as 'to happen' and 'reality' proved to be
unapplicable to this universe  \cite{Har}. Of course, this situation cannot but worry
those physicists to deal with the foundations of QM. For example, Ghirardi \cite{Gh2}
has attached great importance to elaborating the ontological aspects of the
GWRP-model. However, so far there is no consensus in solving this problem.

Where does the root of this problem lie really, at the macroscopic level or no? As is
stated in \cite{Gh1,Gh2,Pe1}, the GWRP-approach retains unchanged the quantum dynamics
of single electrons and atoms. However, in our opinion, in order to keep the
Schr\"odinger equation and terms 'to happen' and 'reality' as well as to construct
eventually a knowable universe, one should discard the current interpretation of the
state (\ref{1}) and solve the problem associated with this state just at the level of
a single electron.

\section{Nonlocal one-particle correlations in the standard model of a 1D completed
scattering} \label{s2}

In solving the Cat paradox, we have to proceed from the fact that the electron-cat
relationship in this paradox is purely {\it causal}; here the cat's fate depends
strongly on the electron's fate. However, the crucial point of the problem is that the
standard model of a 1D completed scattering forbids, in principle, to say that the
electron taking part in this process is {\it either} transmitted {\it or} reflected by
the barrier.

Indeed, for all stages of scattering, this model makes no provision for a separate
description of transmission and reflection, even on the mathematical level. Formally,
the above division on the transmitted and reflected subensembles appears only at the
final stage of scattering, at $t\to\infty$, when the electron's quantum "trajectory"
coincides with the asymptote
$|\Psi_{full}^{end}\rangle=|\Psi_{tr}^{end}\rangle+|\Psi_{ref}^{end}\rangle$, with the
wave packets $|\Psi_{tr}^{end}\rangle$ and $|\Psi_{ref}^{end}\rangle$ occupying the
macroscopically distinct spatial regions. This model implies that the fate of a single
electron to be in this superposition remains indefinite even though there is no
interference between $|\Psi_{tr}^{end}\rangle$ and $|\Psi_{ref}^{end}\rangle$.

This model instructs one to consider the state $|\Psi_{full}^{end}\rangle$ as the
elementary one-electron state and to introduce observables namely for it. However, the
nonphysical character of this instruction becomes obvious when one attempts to
calculate the expectation values for the electron's position $\hat{x}$ and momentum
$\hat{p}$. It is evident that
$\overline{x}_{end}(t)=\langle\Psi_{full}^{end}|\hat{x}|\Psi_{full}^{end}\rangle$ and
$\overline{p}_{end}(t)=\langle\Psi_{full}^{end}|\hat{p}|\Psi_{full}^{end}\rangle$ do
not give the most probable values of these observables.

However, a more strange situation arises for the mean-square deviation
$\overline{(\Delta x)^2}(t)\equiv\langle\Psi_{full}^{end}|(\hat{x}^2-
\overline{x}_{end}^2)|\Psi_{full}^{end}\rangle$ to increase infinitely at
$t\to\infty$. Note, its increase takes place not only because of the transmitted and
reflected wave packets diffuse, but also because of they move away from each other.
This quantity shows that there are nonzero correlations between two events to occur in
the macroscopically distinct regions, separated by the spatial-like interval. As is
seen, this situation reminds that to appear in the EPR-Bell paradox. And, as in the
case of the Bell inequalities, there is no doubt that the above prediction of nonlocal
correlations can be in principle verified experimentally. However, how to explain
nonlocal correlations for this {\it one-particle} process?

The above nonlocal neither-transmitted-nor-reflected quantum one-particle dynamics
admits two scenarios: (\i) either that the electron is a nonlocal combined object to
consist from two macroscopically distinct parts, (\i\i) or that the electron is a
point-like object which can teleport through the spatial region where the probability
density is zero (it separates the transmitted and reflected wave packets). Both the
"explanations" are unacceptable for this spinning particle with the {\it elementary}
electrical charge.

It is often considered that this problem disappears within the statistical
interpretation \cite{Ba1,Ba2} of QM. By its, QM is merely a device for calculating
probabilities, and the amplitudes of probability waves correspond to nothing in the
physical world. However, such attitude to the problem is unsatisfactory too. In our
opinion, just because of ignoring the {\it physical} aspects of CSMDSs this "device",
as it stands, fails to treat them.

The main goal of the paper is to show that the either-transmitted-or-reflected quantum
one-particle dynamics, in a 1D completed scattering, does not at all contradict QM. It
can be supported theoretically and experimentally. Yes, the current model of this
process does not imply such dynamics; any decomposition of the time-dependent
state-vector of a particle, at all stages of scattering, is merely beyond the practice
of the current mathematical formalism of QM. However, as is shown in \cite{Ch1,Ch2},
in reality the Schr\"odinger equation involves such a decomposition.

\section{A new, macrorealistic model of a 1D completed scattering}

\subsection{The superposition principle and continuity equation}

The model \cite{Ch1,Ch2} deals with an electron to impinge, from the left, a symmetric
potential barrier localized in the finite spatial region. Let $\Psi_{full}(x;E)$ be
the wave function to describe the whole ensemble of identical electrons with energy
$E$: to the left of the barrier
\[\fl \Psi_{full}(x;E)=\exp(ikx)+A_{full}^{R}\exp(-ikx);\] to the right -
\[\fl \Psi_{full}(x;E)=A_{full}^{T}exp(ikx);\] here $A_{full}^{R}$ and $A_{full}^{T}$ are
the known complex amplitudes of the reflected and transmitted waves, respectively; $x$
is the particle's coordinate; $k=\sqrt{2mE/\hbar^2}$.

As is shown in \cite{Ch1}, $\Psi_{full}(x;E)$ can be uniquely presented in the form
\begin{eqnarray} \label{3}
\fl \Psi_{full}(x;E)=\Psi_{tr}(x;E)+\Psi_{ref}(x;E)
\end{eqnarray}
where $\Psi_{tr}(x;E)$ and $\Psi_{ref}(x;E)$ are solutions of the Schr\"odinger
equation to obey the boundary conditions (\ref{5}): to the left of the barrier,
\begin{eqnarray} \label{4}
\fl \Psi_{tr}(x;E)=A_{tr}^{In}\exp(ikx)+A_{tr}^{R}\exp(-ikx),\nonumber\\
\fl \Psi_{ref}(x;E)=A_{ref}^{In}\exp(ikx)+A_{ref}^{R}\exp(-ikx);
\end{eqnarray}
\begin{eqnarray} \label{5}
\fl A_{tr}^{R}=0,\ooo A_{ref}^{R}=A_{full}^{R},\ooo A_{tr}^{In}+A_{ref}^{In}=1,\ooo
|A_{tr}^{In}|=|A_{full}^{T}|,\ooo |A_{ref}^{In}|=|A_{full}^{R}|.
\end{eqnarray}

Note, there are two sets of the amplitudes $A_{tr}^{In}$ and $A_{ref}^{In}$ to satisfy
the boundary conditions (\ref{5}). One of them leads to the wave function
$\Psi_{ref}(x;E)$ to be even, with respect to the midpoint $x_c$ of the region of the
symmetric potential barrier. Another leads to an odd function. We choose the latter.
In this case, $\Psi_{ref}(x_c;E)=0$ for any value of $E$. And, at any value of $t$,
wave packets formed from the odd solutions are equal to zero at this point, too. This
means that electrons to impinge the barrier from the left do not enter the region
$x>x_c$.

However, we have to stress that both functions, $\Psi_{tr}(x;E)$ and
$\Psi_{ref}(x;E)$, contain the terms to describe electrons impinging the barrier from
the right, which disappear in the superposition (\ref{3}) due to interference. As a
result, in this superposition, electrons impinging the barrier from the left and then
being reflected (transmitted) by its are described by the function $\psi_{ref}(x;E)$
($\psi_{tr}(x;E)$) where
\begin{eqnarray*}
\fl \psi_{ref}(x;E)\equiv \Psi_{ref}(x;E),\ooa \psi_{tr}(x;E)\equiv
\Psi_{tr}(x;E),\ooa
x\leq x_c;\nonumber\\
\fl \psi_{ref}(x;E)\equiv 0,\ooa \psi_{tr}(x;E)\equiv \Psi_{full}(x;E), \ooa x>x_c.
\end{eqnarray*}
Now $\Psi_{full}(x;E)=\Psi_{tr}(x;E)+\Psi_{ref}(x;E)\equiv
\psi_{tr}(x;E)+\psi_{ref}(x;E)$.

As is seen, the first derivatives on $x$ of the functions $\psi_{tr}(x;E)$ and
$\psi_{ref}(x;E)$ are discontinuous at the point $x_c$. This results from the fact
that only the sum of these functions obeys the Schr\"odinger equation. The either
function obeys the continuity equation. The same holds for all wave packets formed
from these functions.

Let $\Psi_{full}(x,t)$ be a solution of the time-dependent Schr\"odinger equation for
a given initial condition. Let also $\Psi_{tr}(x,t)$ and $\Psi_{ref}(x,t)$ be the
corresponding solutions formed from $\Psi_{tr}(x;E)$ and $\Psi_{ref}(x;E)$,
respectively. Besides, let $\psi_{tr}(x,t)$ and $\psi_{ref}(x,t)$ be the corresponding
wave packets formed from $\psi_{tr}(x;E)$ and $\psi_{ref}(x;E)$. Then we have
\begin{eqnarray} \label{6}
\fl \Psi_{full}(x,t)=\Psi_{tr}(x,t)+\Psi_{ref}(x,t)\equiv
\psi_{tr}(x,t)+\psi_{ref}(x,t)
\end{eqnarray}

By \cite{Ch1}, namely $\psi_{tr}(x,t)$ and $\psi_{ref}(x,t)$ describe, at all stages
of scattering, the motion of the (to-be-)transmitted and (to-be-)reflected
subensembles. Both, $\psi_{tr}(x,t)$ and $\psi_{ref}(x,t)$, are solutions to the {\it
real} continuity equation, and their sum obeys the {\it complex} Schr\"odinger
equation. Hence $\psi_{tr}(x,t)+\psi_{ref}(x,t)$, unlike
$\Psi_{tr}(x,t)+\Psi_{ref}(x,t)$, is the superposition of probability waves to {\it
interact} with each other, excepting the limiting case when $t\to \infty$. That is, in
fact, we have presented the time evolution of a closed ensemble of scattering
electrons as a coherent evolution of the open (to-be-)transmitted and
(to-be-)reflected subensembles of electrons (it is relevant here to point to the
recent paper \cite{Sla}). This results from the fact that, for a given
semi-transparent potential, the transmission and reflection are inseparable
sub-processes.

Of importance is that for any value of $t$ the scalar product
$\langle\psi_{tr}(x,t)|\psi_{ref}(x,t)\rangle$ is a purely imagine value to diminish
at $t\to\infty$. Despite interference between $\psi_{tr}$ and $\psi_{ref}$,
\begin{eqnarray} \label{7}
\fl \langle\Psi_{full}(x,t)|\Psi_{full}(x,t)\rangle =T+R=1
\end{eqnarray}
where $T=\langle\psi_{tr}(x,t)|\psi_{tr}(x,t)\rangle=const$,
$R=\langle\psi_{ref}(x,t)|\psi_{ref}(x,t)\rangle=const$; $T$ and $R$ are the
transmission and reflection coefficients, respectively. That is, in the case of the
CSMDS quantum probabilities behave as classical, Kolmogorovian probabilities.

\subsection{Non-invasive measurements for the sub-processes}

So, by the model, a 1D completed scattering is a combined process to consist from two
coherently evolved sub-processes, transmission and reflection. That is, at any instant
of time the state of the ensemble of scattering electrons is a CSMDS. Hence, to
observe the scattering of the wave packet on a 1D potential barrier means, in fact, to
observe the interference pattern formed by the sub-processes.

However, the main peculiarity of any combined quantum process is that it also implies
performing experiments for testing the individual properties of sub-processes to form
it. In \cite{Ch2}, both for transmission and reflection we have introduced the dwell
time to give the time spent by an electron in the barrier region. For an electron with
a given energy, this quantity is defined via the probability current density and
probability density. As regards the time-dependent transmission and reflection, the
either is described by the Larmor time to represent the average value of the dwell
time, which can be measured by means of the non-invasive, Larmor-clock procedure.

As is known \cite{But}, this procedure implies switching on an infinitesimal magnetic
field in the barrier region. Then the angle of the Larmor precession of the average
electron's spin is measured separately for the transmitted and reflected subensembles,
well after the scattering event. That is, in this procedure the average electron's
spin serves as a clock-pointer to "remember" the time spent by an electron in the
barrier region. It is evident that all measurements performed on transmitted  or
reflected electrons do not influence those performed for alternative sub-process, for
all measurement are carried out when there is no interference and interaction between
$\psi_{tr}(x,t)$ and $\psi_{ref}(x,t)$.

Note, unlike the previous definition \cite{But} of this characteristic time, ours do
not predict the Hartman effect whose nature remains unclear up to the present
\cite{Win}.

\section{Discussion and conclusion}

So, the cat's fate in the above version of the Cat paradox is quite definite, as the
electron's fate is definite. This closed one-electron process is a combined one to
consist from two open sub-processes, transmission and reflection, evolved coherently.
In a single experiment, a single electron to be a point-like object is {\it either}
transmitted {\it or} reflected by the potential barrier. The wave function to describe
the whole ensemble of scattering electrons is the superposition of those to describe
its (to-be-)transmitted and (to-be-)reflected parts. Being in the quantum
superposition, either sub-process does not lose however its individual properties
which can be experimentally examined, in any spatial interval, by means of the
non-invasive Larmor-clock procedure.

In fact, our model extends the validity of the Schr\"odinger equation onto the
macroscopic scales. It says that the PMRs must be considered as a part of basic
principles of QM. Within this vision of QM, the wave-particle duality implies that, at
the atomic scales, a single electron behaves (unpredictably) as a point-like object,
while the electron's ensemble behaves (deterministically) as a wave. The one-electron
wave function describes a single electron in the (strictly speaking, infinite) set of
identical experiments (i.e., the electron's ensemble), rather than a single electron
in a single experiment. That is, QM is a complete theory of ensembles. To explain the
quantum dynamics of a single electron is a prerogative of a future sub-quantum theory,
which should appeal to the electron's structure, i.e., to the scales of order
$10^{-13}$ cm.

Any pure time-dependent state to involve at least two macroscopically distinct
alternatives for a particle must have a special status in the macrorealistic QM:

(1) It should be considered as a pure combined state - the intermediate link between a
pure elementary one (indecomposable into macroscopically distinct parts) and
statistical mixture.

(2) QM must provide the presentation of a pure time-dependent combined state as a
CSMDS, for all moments of time. Any combined quantum one-particle process to consist
from $N$ coherently evolved elementary sub-processes should be considered as the
counterpart to $N$ classical one-particle trajectories.

(3) In the case of a pure combined process, all observables can be introduced only for
its elementary sub-processes. Any combined process needs two types of measurements:
(\i) those for observing the interference pattern resulting from a joint action of all
the coherently evolved sub-processes, and (\i\i) non-demolishing 'which-way'
measurements for inspecting the individual properties of these sub-processes (they are
performed at the stages when the sub-processes do not interfere with each other).

Note, a combined process itself cannot be endowed with observables. Disregarding this
rule, as in the standard model of a 1D completed scattering (Section \ref{s2}), leads
inevitable to nonlocality. The interference pattern for this process can be properly
interpreted only with taking into account that it is formed by two sub-processes.

\end{document}